\documentclass[aps,amssymb,floatfix,prd,amsmath,preprintnumbers]{revtex4}
\usepackage{amsfonts}
\usepackage{amssymb}
\usepackage{amsmath}
\usepackage{graphicx}  
\usepackage{mathtools}

\DeclarePairedDelimiter\ket{\lvert}{\rangle}
\DeclarePairedDelimiterX\braket[2]{\langle}{\rangle}{#1 \delimsize\vert #2}
\usepackage{float}
\usepackage{dcolumn}   
\usepackage{bm}
\begin{document}

\title{\bf  Modelling Casimir Wormholes in  Extended Gravity}

\author{Sunil Kumar Tripathy\footnote{E-mail:tripathy\_ sunil@rediffmail.com}
}
\affiliation{Department of Physics, Indira Gandhi Institute of Technology, Sarang, Dhenkanal, Odisha-759146, India }

\begin{abstract}
A model of traversable wormhole in an extended gravity theory has been proposed. The Casimir effect and Generalised Uncertainty Principle (GUP) arising out of the concept of minimal length have been considered to obtain the shape function, radial and tangential pressure of the wormhole. The effects of the GUP parameter and the parameter of the extended gravity theory on the wormhole properties have been discussed. The energy condition violation by the exotic matter of Casimir wormhole have been shown. Also we have calculated the exotic matter content of the Casimir wormhole that violates the null energy condition.
\end{abstract}
\maketitle
\section{Introduction} 

Wormholes are the hypothetical bridges connecting two asymptotic regions of the same spacetime  and provide a short cut path that can be traversable through a minimal surface area called wormhole throat. Einstein and Rosen, proposed the existence of such hypothetical passages called Einstein-Rosen bridges in General Relativity (GR) \cite{Einstein35}. Morris and Thorne proposed traversable wormholes as the solutions of Einstein field equations that contain exotic matter with negative energy density \cite{Morris88}. The stability of the traversable wormhole depends on its exotic matter content. It is obvious that the exotic matter of wormholes violates the null energy condition (NEC). There have been many attempts in literature to avoid or to minimize the NEC violation. One such example is the model of Visser \cite{Visser95}. Visser constructed wormhole models in such a manner that traversable path should not fall within the area of exotic matter. Classically, it is not possible to have enough negative energy density and therefore no traversable wormholes can exist. However, quantum mechanical concepts provide us opportunity to have negative energy sources through the realisation of vacuum fluctuation of  quantum fields. Also, Marolf and Polchinski have shown that wormholes are associated with some sort of quantum entanglement \cite{Marolf2013}.

Till date, no wormholes have been detected or even  there is no trace of the consequences of traversable wormholes.  Recently, Garattini \cite{Garattini19} has proposed a model of static traversable wormhole exploring the negative energy density due to the Casimir effect and examined the consequences of the constraint imposed by a quantum weak energy condition. In fact, the Casimir effect has a strong dependence on the geometry of boundaries and the Casimir energy represents the only artificial source of exotic matter realizable in laboratory.  The quantum field of the vacuum between the two uncharged plane parallel plates are distorted and leads to a negative energy density which may well be considered as a suitable source for traversable wormholes. Another important aspect of quantum mechanics developed in recent times is the concept of minimal length of the order of Planck length. The minimal length scale limits the resolution of small distances in spacetime. This arises naturally in quantum gravity theories with a minimum uncertainty in position. This concept of minimal length obviously demands for a change in the position-momentum uncertainty relation. The generalised uncertainty principle (GUP) redefines the negative Casimir energy density. Of course, this redefined Casimir energy density depends on the specific construction of the maximally localized quantum states. In view of this, it is interesting to incorporate the effect of GUP in modelling traversable wormholes. Jusufi et al. have investigated the effect of the GUP in the Casimir wormholes where the source of exotic matter is nothing but the negative Casimir energy density \cite{Jusufi19}. Javed et al. \cite{Javed20} have calculated the weak deflection angle of light from Casimir wormholes in the weak field limit.

In the present work, we have explored the effect of the GUP  on the Casmir wormholes in the framework of an extended gravity theory. It is needless to mention here that, recent observations suggest that the universe is not only expanding but the expansion is accelerating at the present epoch. In the purview of GR, this late time cosmic speed up issue can be attributed to an exotic dark energy form with a negative pressure. However, different modified theories of gravity  have been proposed in recent times without the need of any such dynamical degrees of freedom \cite{Caroll2004, Nojiri2007, Nojiri2005, Linder2010, Myrza2011, Harko2011}. In these modified theories, the geometrical action is modified with or without the inclusion of any matter form. The material contributions within the geometrical action is usually motivated from the quantum effects such as the particle production. The extended theories of gravity having a material contribution in the geometrical action have been quite successful in addressing many issues in cosmology and astrophysics. Of late, modified theories of gravity are becoming popular for modelling wormholes \cite{Yousaf17, Zubair15, Moraes17, Elizalde19, Elizalde19a, Elizalde19b, Rahaman15, Godani18, Godani20}.

The organisation of the paper is as follows: Section II contains a brief review of the extended gravity theory. In Section-III, a model of traversable wormholes is presented in the extended gravity theory. Wormhole model with the incorporation of Casimir effect  is discussed in Section-IV.  In this section, we have considered the negative energy density due to Casimir effect to be the source of exotic matter in traversable wormholes and studied the effect of the extended gravity parameter on the shape function and other properties. A brief review of the quantum mechanics of the generalised uncertainty principle is  presented in Section-V. The shape function, radial and tangential pressure for Casimir wormholes with the GUP correction are discussed in Section-VI. Two different types of construction of the maximally localized states have been considered in the present work to model the GUP corrected Casimir wormholes. The effect of the minimal uncertainty parameter of GUP and the extended gravity parameter on the wormhole properties are investigated. The energy condition violation along with the amount of exotic matter content in the Casimir wormhole have been calculated in Section VII.  The results of the present work are summarised in Section-VIII.

\section{A brief review of the  Formalism}
We consider the action for a geometrically modified theory of gravity as 
\begin{equation} \label{eq:1}
S=\int d^4x\sqrt{-g}\left[\frac{1}{16\pi} f(R,T)+ \mathcal{L}_m \right],
\end{equation}
where $f(R,T)$ is an arbitrary function of the Ricci scalar $R$ and the trace $T$ of the energy-momentum tensor. $\mathcal{L}_m$ is the matter Lagrangian.  The action reduces to that of GR for $f(R,T)=R$.  We use the natural system of unit: $G=c=\hbar= 1$; $G$ is the Newtonian gravitational constant, $c$ is  the speed of light in vacuum and  $\hbar$  is the reduced Planck constant.

For minimal matter-geometry coupling, we may consider $f(R,T)=f_1(R)+f_2(T)$. A variation of the action with respect to the metric $g_{\mu\nu}$, yields the field equations as 

\begin{equation} \label{eq:3}
R_{\mu\nu}-\frac{1}{2}f^{-1}_{1,R} (R)f_1(R)g_{\mu\nu}=f^{-1}_{1,R}(R)\left[\left(\nabla_{\mu} \nabla_{\nu}-g_{\mu\nu}\Box\right)f_{1,R}(R)+\left[8\pi +f_{2,T}(T)\right]T_{\mu\nu}+\left[f_{2,T}(T)p+\frac{1}{2}f_2(T)\right]g_{\mu\nu}\right].
\end{equation}

Here we have used the shorthand notations
\begin{equation}\label{eq:4}
f_{1,R} (R)\equiv \frac{\partial f_1(R)}{\partial R},~~~~~~~~~ f_{2,T} (T)\equiv \frac{\partial f_2(T)}{\partial T}, ~~~~~~~~~ f^{-1}_{1,R} (R) \equiv \frac{1}{f_{1,R} (R)}.
\end{equation}
Following Harko et al. \cite{Harko2011}, we have assumed the matter Lagrangian as $\mathcal{L}_m=-p$. The energy-momentum tensor $T_{\mu\nu}$ can be obtained from the matter Lagrangian:
\begin{equation}\label{eq:5}
T_{\mu\nu}=-\frac{2}{\sqrt{-g}}\frac{\delta\left(\sqrt{-g}\mathcal{L}_m\right)}{\delta g^{\mu\nu}}.
\end{equation}

For a simple choice $f_1(R)=R$, the field equation reduces to
\begin{equation}\label{eq:6}
G_{\mu\nu}= \left[8\pi +f_{2,T}(T)\right]T_{\mu\nu}+\left[f_{2,T}(T)p+\frac{1}{2}f_2(T)\right]g_{\mu\nu}.
\end{equation}
We can write the field equation as
\begin{equation}\label{eq:7}
G_{\mu\nu}= \kappa_{T}\left[T_{\mu\nu}+ T^{int}_{\mu\nu}\right].
\end{equation}
where $G_{\mu\nu}$ is the Einstein tensor. $\kappa_{T}= 8\pi +f_{2,T}(T)$ is the redefined Einstein constant which may depend on the trace of the energy momentum tensor.  The additional interactive energy-momentum tensor due to the geometrical modification through a minimal coupling with matter is given by 

\begin{equation}\label{eq:8}
T^{int}_{\mu\nu}=\left[ \frac{f_{2,T}(T)p+\frac{1}{2}f_2(T)}{8\pi +f_{2,T}(T)}\right]g_{\mu\nu}.
\end{equation}
This term is due to the additional $f_2(T)$ term in the geometrical action. A viable model may be constructed with a suitable choice of $f_2(T)$ which may be confronted with recent observations.

For the sake of brevity, in the present work, we consider a linear functional 

\begin{equation}\label{eq:9}
\frac{1}{2}f_2(T)=\lambda T,
\end{equation}
so that $\kappa_T = 8\pi+2\lambda$ and $T^{int}_{\mu\nu} = \frac{g_{\mu\nu}}{\kappa_T}\left[\left(2p+T\right)\lambda\right]$.  $\kappa_T$ reduces to the usual Einstein constant in the limit $\lambda \rightarrow 0$ and consequently the interactive term $T^{int}_{\mu\nu}$ vanishes. In other words, the GR can be recovered from the extended gravity in the limit $\lambda \rightarrow 0$.  Since, GR without any dynamical degrees of freedom fails to explain the late time cosmic speed up phenomenon, it is wise to consider a simple geometrical modification to achieve the necessary cosmic acceleration. In this theory, the additional interactive term in the field equation takes care of the cosmic speed up issue. One can note that, similar choices of the functional $f(R,T)$ have been widely used in literature for investigations of many issues in cosmology and astrophysics \cite{Harko2011, Shabani2014, Das2017,  Moraes2017, Deb2018, Yousaf2018, Sharif2019, Mishra18a, Mishra18b, Yousaf16, Sahu2017, SKT19,PKS2018, SKT20, Biswas2020}. Moreover, recently Ordines and Calson  have tried to constrain this coupling parameter $\lambda$ from the observational data on earth's atmosphere \cite{Ordines2019}.

\section{Wormholes in extended gravity}
A static and spherically symmetric Morris-Thorne traversable wormhole is modelled through the metric given in Schwarzchild coordinates \cite{Morris88}
\begin{equation}
ds^2=-e^{2\Phi}dt^2+\frac{1}{\psi(r)}dr^2+r^2d\Omega^2,
\end{equation}
where $\Phi=\Phi(r)$ is the redshift function and $\psi(r)=1-\frac{b(r)}{r}$. Here $d\Omega^2=d\theta^2+sin^2\theta d\phi^2$ is the surface element. $b(r)$ is the shape function defining the geometry of the wormhole. The redshift function and the shape function must satisfy some of the conditions to provide a meaningful wormhole geometry. The redshift function $e^{2\Phi}$ must be finite and non-vanishing to avoid the formation of event horizon. The shape function $b(r)$ reduces to the size of the wormhole throat radius $r_0$ at the throat i.e $b(r_0)=r_0$ and should obey the flare out condition $\frac{b(r)-rb^{\prime}}{b^2(r)} >0$. If we evaluate this flare out condition at the wormhole throat, the condition reduces to $b^{\prime}(r_0)=\frac{db(r)}{dr}|_{r=r_0}<1$. 

The exotic matter content of the wormhole is described through the anisotropic fluid
\begin{equation}
T_{\mu\nu}= \left(\rho+p_t\right)u_{\mu}u_{\nu}+p_tg_{\mu\nu}+\left(p_r-p_t\right)x_{\mu}x_{\nu},
\end{equation}
where $u_{\mu}$ and $x_{\mu}$ are the four velocity vectors along the transverse and radial directions. $u_{\mu}$ and $x_{\mu}$ satisfy the relations $u^{\mu}u_{\mu}=-1$ and $x^{\mu}x_{\mu}=1$. $\rho$ is the energy density. $p_t$ and $p_r$ are the tangential and radial components of the pressure. The trace of the energy momentum tensor becomes $T=-\rho+p_r+2p_t$.

For a tideless wormhole, we have $\Phi=0$ and therefore the field equations for the wormhole metric in the extended gravity can be obtained as

\begin{eqnarray}
\rho(r) &=& \kappa^{\prime}_T\frac{b^{\prime}(r)}{r^2},\label{eq:15}\\
p_r(r)  &=& -\kappa^{\prime}_T\frac{b(r)}{r^3},\\
p_t(r)  &=& \kappa^{\prime}_T\frac{b-b^{\prime}r}{2r^3},
\end{eqnarray}
where $\kappa^{\prime}_T=\frac{1}{\kappa_T}$. For GR, we have, $\kappa_T=8\pi$ and consequently the above wormhole field equations reduce to

\begin{eqnarray}
\rho(r) &=&\frac{1}{8\pi}\frac{b^{\prime}(r)}{r^2},\\
p_r(r)  &=& -\frac{1}{8\pi}\frac{b(r)}{r^3},\\
p_t(r)  &=& \frac{1}{8\pi}\frac{b-b^{\prime}r}{2r^3}.
\end{eqnarray}
\section{Casimir Wormholes}
The Casimir effect involves the manifestation of an attractive force between a closely placed pair of neutral, parallel uncharged conducting plates in vacuum.  This interaction is caused by the disturbance of the vacuum of the electromagnetic field and can be associated with the zero-point energy of a quantum electrodynamics (QED) vacuum distorted by the plates. Such an effect is predicted theoretically by H. Casimir in 1948 \cite{Casimir48} but observed experimentally later \cite{Sparnaay57, Bressi2002}. Casimir effect is a pure quantum effect since the ground state of quantum electrodynamics or the vacuum that causes the the uncharged plane plates to attract each other. In fact, the Casimir effect has a strong dependence on the geometry of boundaries and the Casimir energy represents the only artificial source of exotic matter realizable in laboratory \cite{Garattini19}. In general, exotic matter violates the energy condition particularly the null energy condition. In principle, it is reasonable to consider Casimir effect in traversable wormholes which contains exotic matter violating NEC. With such an idea, Garattini has proposed a model of traversable wormholes using the equation of state arising out of the Casimir effect. Such wormholes are named as Casimir wormholes \cite{Garattini19}.

According to the Casimir effect, the attractive force between the plates arise because of the renormalized negative energy
\begin{equation}
E(a)=-\frac{\pi^2}{720}\frac{S}{a^3},
\end{equation}
where $a$ is the distance of separation between the plates and $S$ is the surface area of the plates. One can note that the Casimir energy is lowered if we move the two plates closer. Consequently, the energy density becomes
\begin{equation}
\rho(a)=-\frac{\pi^2}{720}\frac{1}{a^4}.\label{eq:rho}
\end{equation}

The pressure can be obtained from the renormalized negative energy as
\begin{equation}
p(a)=-\frac{1}{S}\frac{dE(a)}{da}=-\frac{\pi^2}{240}\frac{1}{a^4}.\label{eq:p}
\end{equation}
The above expressions of energy density and pressure lead to an equation of state (EoS) $p=\omega\rho$ with the EoS parameter $\omega=3$. The Casimir force is defined as the surface area times the pressure and can be given by $F=-\frac{\pi^2S}{240}\frac{1}{a^4}$. This force is a negative quantity and is therefore attractive. 

It is believed that, traversable wormholes require exotic matter content violating the NEC for their existence.   The exotic matter stabilizes a wormhole by allowing it to collapse slowly and may remain open for a long time \cite{Butcher2014}. In his work, Butcher has shown that, the stability of a traversable wormhole can be possible if the wormhole has a large throat as compared to the Planck scale \cite{Butcher2014}. In a recent work, Garattini \cite{Garattini2020} has discussed traversable wormholes sourced by Casimir energy by assuming a fixed plate separation distance. With a fixed plate separation, Garattini showed that, stable traversable wormholes are possible but with huge throat compared to the Planckian scale. In the present work, we assume that, the exotic matter of our wormhole be associated with the Casimir energy density given in \eqref{eq:rho} with a replacement of the plate separation distance $a$ by the radial coordinate $r$. Such a choice favours wormhole throat of Planckian scale \cite{Garattini19}.

Integrating eq.\eqref{eq:15} using the Casimir energy density in eq. \eqref{eq:rho} and imposing the throat condition  $b(r_0)=r_0$, we obtain the shape function as
\begin{equation}
b(r)=r_0+b_1\left(\frac{1}{r}-\frac{1}{r_0}\right),
\end{equation}
where $b_1=\frac{\pi^2\kappa_T}{720}$.

Since, $b^{\prime}=-\frac{b_1}{r^2}$, the flaring out condition calculated at the wormhole throat becomes $-\frac{b_1}{r_0^2}<0$. This condition implies that $b_1$ should be positive and consequently we may constrain the parameter $\lambda$ in the range $\lambda >-4\pi$. 

\begin{figure}[t!]
\begin{center}
\includegraphics[scale=0.4]{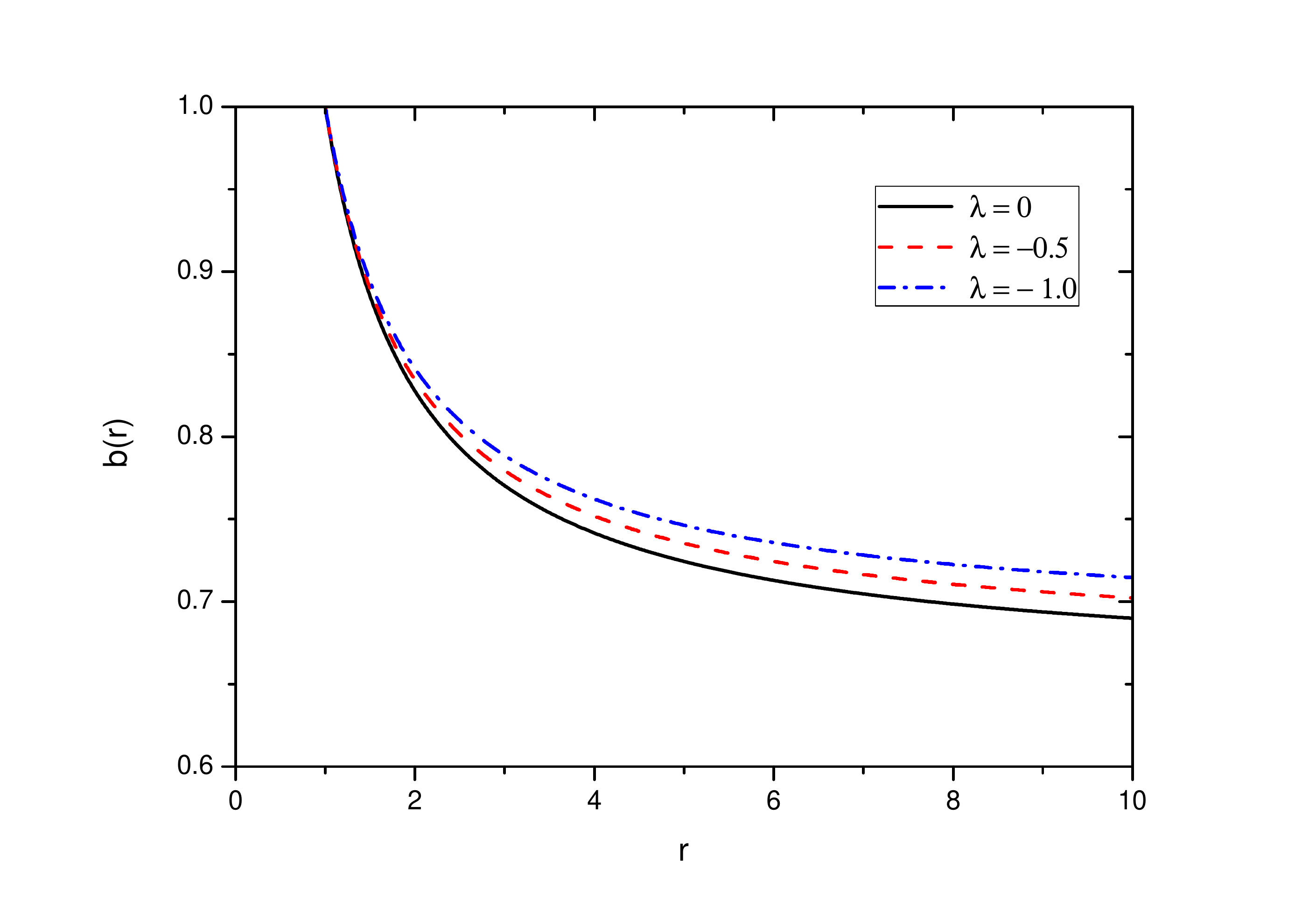}
\caption{Shape function $b(r)$ of the Casimir wormhole for three different values of $\lambda$. The throat radius is considered as $r_0=1$. In the figure $\lambda=0$ corresponds to that in GR. }
\end{center}
\end{figure}

In Figure 1, we have plotted the shape function for the Casimir wormhole in the extended gravity theory for some possible values of the parameter $\lambda$ namely $\lambda=0, -0.5$ and $-1$. All of these choices of $\lambda$ are within the constrained values $\lambda >-4\pi $.  For $\lambda=0$, we recover the behaviour in GR. In the figure, the throat radius is taken as $r_0=1$ so that the radial distance is measured in the unit of the throat radius. In the present work, we have assumed an asymptotically flat wormhole metric and as expected for all the assumed values of $\lambda$, the shape function asymptotically assumes the value of $b(r\rightarrow \infty)=r_0-\frac{b_1}{r_0}$. The effect of the parameter $\lambda$ on the shape function is quite visible beyond the wormhole throat. For a given radial distance, a decrease in the value of $\lambda$ increases the value of the shape of function $b(r)$. The rate of increment in $b(r)$ with respect to $\lambda$, increases as we move away from the throat. Also, the asymptotic value $b(r\rightarrow \infty)$ increases with a decrease in $\lambda$. The quantity  $b^{\prime}=-\frac{b_1}{r^2}$ vanishes asymptotically signifying the asymptotically flat wormhole metric. Also we have 
\begin{equation}
\lim_{r\rightarrow\infty} \frac{b(r)}{r}=0.
\end{equation}

The radial and tangential pressure for the Casimir wormhole in the extended gravity become
\begin{eqnarray}
p_r(r) &=& -\kappa^{\prime}_T\frac{1}{r_0r^4} \left[rr_0^2+b_1\left(r_0-r\right)\right],\\
p_t(r) &=& \frac{1}{2}\kappa^{\prime}_T\frac{1}{r_0r^4} \left[rr_0^2+b_1\left(2r_0-r\right)\right].
\end{eqnarray} 

Let us now define a radial EoS parameter $\omega_r(r)=\frac{p_r(r)}{\rho(r)}$ which can be obtained as
\begin{equation}
\omega_r(r)=\frac{rr_0^2+b_1\left(r_0-r\right)}{b_1r_0}.
\end{equation}

Similarly for a scenario with the equation of state $p_t(r)=\omega_t(r)\rho(r)$, we can have

\begin{equation}
\omega_t(r)=-\frac{rr_0^2+b_1\left(2r_0-r\right)}{2b_1r_0}=-\frac{1}{2}\left[1+\omega_r(r)\right].
\end{equation}
The behaviour of the EoS parameters $\omega_r(r)$ and $\omega_t(r)$ are shown in Figures 2(a) and 2(b) respectively. The radial equation of state parameter increases with radial distance whereas the tangential equation of state parameter decreases with the radial distance. At the throat, the radial and tangential EoS parameters reduce to $\omega_r(r_0)=\frac{r_0^2}{b_1}$ and $\omega_t(r_0)=\frac{r_0^2}{2b_1}+\frac{1}{2}$ respectively. At a given radial distance, while $\omega_r(r)$ increases with a decrease in $\lambda$, $\omega_t(r)$ decreases. The radial and lateral pressure on throat become $p_r(r_0)=-\frac{\kappa^{\prime}_T}{r_0^2}$ and $p_t(r_0)=\frac{\kappa^{\prime}_T}{r_0^2}\left[\frac{1}{2}+\frac{b_1}{2r_0^2}\right]$. From the radial and tangential pressure we can have an idea of the pressure anisotropy at the throat as $|\frac{p_t(r_0)}{p_r(r_0)}|=|\frac{\omega_t(r_0)}{\omega_r(r_0)}|=\frac{1}{2}+\frac{b_1}{2r_0^2}$. This anisotropy increases with a decrease in the throat radius. Since, $b_1=\frac{\pi^2}{720}(8\pi+2\lambda)$ decreases with the decrease in the value of the parameter $\lambda$, the pressure anisotropy decreases for a given wormhole throat.

With the assumption of Casimir energy density, the metric for the wormhole in extended gravity theory reads as
\begin{equation}
ds^2=-dt^2+\left[\frac{r-r_0-b_1\left(\frac{1}{r}-\frac{1}{r_0}\right)}{r}\right]^{-1}dr^2+r^2d\Omega^2.
\end{equation}

\begin{figure}[h!]
\begin{center}
\includegraphics[scale=0.4]{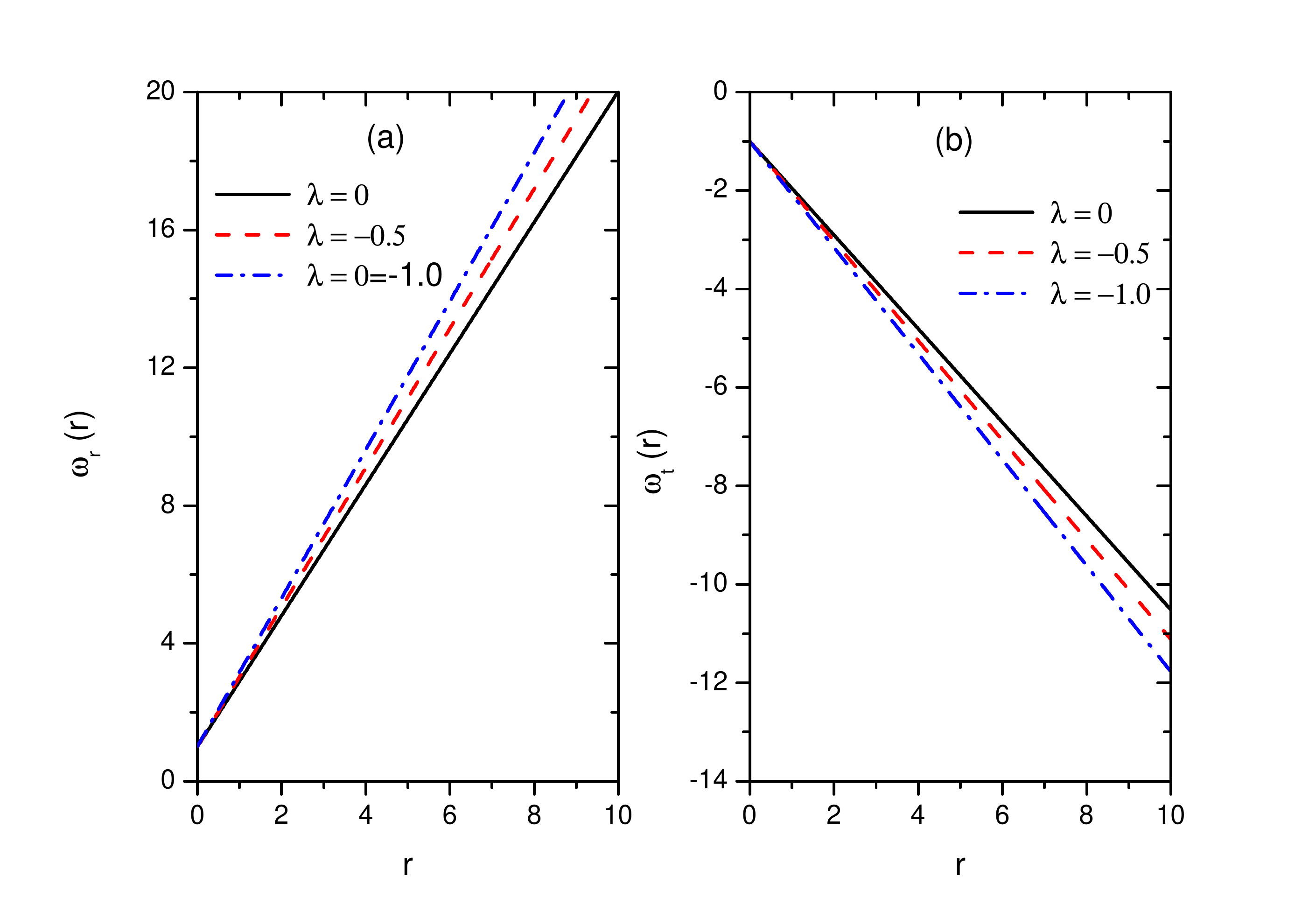}
\caption{(a) Radial equation of state parameter $\omega_r(r)$ of the Casimir wormhole for three different values of $\lambda$. (b) Transverse EoS parameter $\omega_t(r)$ of the Casimir wormhole for three different values of $\lambda$. The throat radius is considered as $r_0=1$. In the figure $\lambda=0$ corresponds to that in GR. }
\end{center}
\end{figure}

\section{Quantum Mechanics of Generalized Uncertainty Principle}

Various theories of quantum gravity such as the string theory \cite{Veneziano89, Konishi90, Maggiore93, Witten96}, loop quantum gravity\cite{Garay95}, non commutative spacetime \cite{Maggiore93a}, doubly special relativity \cite{Magueijo02, Cortes05} and others \cite{Kempf95, Nozari12} imply the existence of a minimal length scale of the order of Planck length $l_p=\sqrt{\frac{G\hbar}{c^3}}\simeq 10^{-35}m$  that limits the resolution of small distances in the spacetime \cite{Pedram2012}. This arises naturally in quantum gravity theories in the form of an effective minimal uncertainty in positions $\triangle x_0 >0$. In string theory, there is a fundamental length scale that determines the typical spacetime extension $l_s=\sqrt{\alpha^{\prime}}$ of a fundamental string with string tension $\frac{\hbar c}{\alpha^{\prime}}$. Moreover, $l_p\sim l_s$  and one can not improve the spatial resolution below this characteristic length i.e.
\begin{equation}
\triangle x\geq l_p\sim l_s.
\end{equation}
This existence of minimal length demands a correction in the position-momentum uncertainty relation in quantum mechanics. At least in one dimension, the generalised uncertainty principle(GUP) can be written as
\begin{equation}
\triangle x\triangle p \geq \frac{1}{2}\left[1+\beta\left(\triangle p\right)^2\right],
\end{equation}
with $\beta =\alpha^{\prime}$. Here we have assumed $\hbar=1$. This GUP predicts a finite non-zero minimum uncertainty in position $\triangle x_0 = \frac{3\sqrt{3}}{4}\sqrt{\beta}$. The commutator of the position and momentum operators now changes from $\left[\hat{x}, \hat{p}\right]=i$, ($i=\sqrt{-1}$) to the form \cite{Pedram2012, Pedram2012a}

\begin{equation}
\left[\hat{x}, \hat{p}\right] = \frac{i}{1-\beta\hat{p}^2}.\label{eq:29}
\end{equation}
With an expansion of the right side of the above equation \eqref{eq:29}, the commutator can be written as 
\begin{equation}
\left[\hat{x}, \hat{p}\right]=i\left(1+\beta\hat{p}^2+\cdots\right).\label{eq:UP}
\end{equation}

Interestingly, the matrix elements of the eigenstates of the position operator $\left\langle x|\psi\right\rangle$ do not have a straight physical interpretation. This is because, the eigenstates do not corresponds to real physical states. In view of this, quasi-position representation is adopted which projects the states onto a set of maximally localized states $\ket{\psi^{ML}}$.
The maximally localized states in momentum representation are given by
\begin{equation}
\psi^{ML}=\frac{1}{\left(2\pi\right)^{3/2}}~\Omega(p)~\text{exp}^{-i\left[\textbf{k}\cdot \textbf {r}-\text{w}(p)t\right]},
\end{equation}
which are the solution of the equation
\begin{equation}
\left[\hat{x}-\left\langle x\right\rangle +\frac{\left\langle \left[\hat{x}, \hat{p}\right]\right\rangle}{2\left(\triangle p\right)^2}\left(\hat{p}-\left\langle p\right\rangle\right)\right]\ket{\psi}=0.
\end{equation}
The minimal length corrected commutation relation in eq. \eqref{eq:UP} can be generalized to $n$ dimension as \cite{Frassino12}
\begin{equation}
\left[\hat{x}_i, \hat{p}_j\right]=i\left[f(\hat{p}^2)\delta_{ij}+g(\hat{p}^2)\hat{p}_i\hat{p}_i\right], ~~~~~~~~~ i,j=1,\cdots, n,
\end{equation}
where $f(\hat{p}^2)$ and $g(\hat{p}^2)$ are the generic functions and can be obtained from translational and rotational invariance of the generalized commutation relation. Different possible choices of the generic functions lead to different models and consequent maximally localized states. In the present work, we will discuss two specific choices of the generic functions that lead to two different models. One that is proposed by Kempf, Mangano and Mann (KMM)\cite{Kempf95} and the other  that has been proposed by Detournay, Gabriel and Spindel (DGS) \cite{Detournay02}. 

The generic functions in these models are given by \cite{Frassino12, Detournay02, Kempf97}
\begin{eqnarray}
f(\hat{p}^2) &=& \frac{\beta \hat{p}^2}{\sqrt{1+2\beta \hat{p}^2}-1},\\
g(\hat{p}^2) &=& \beta.
\end{eqnarray}

The KMM construction of the maximally localized states involves
\begin{eqnarray}
k_i &=& \left(\frac{\sqrt{1+2\beta p^2}-1}{\beta p^2}\right)p_i,\\
\text{w}(p) &=& \left(\frac{\sqrt{1+2\beta p^2}-1}{\beta p^2}\right)p,\\
\Omega(p) &=& \left(\frac{\sqrt{1+2\beta p^2}-1}{\beta p^2}\right)^{\alpha/2},
\end{eqnarray}
where $\alpha=1+\sqrt{1+\frac{n}{2}}$, $n$ being the number of spatial dimensions. 

The DGS construction of the maximally localized states involves
\begin{eqnarray}
k_i &=& \left(\frac{\sqrt{1+2\beta p^2}-1}{\beta p^2}\right)p_i,\\
\text{w}(p) &=& \left(\frac{\sqrt{1+2\beta p^2}-1}{\beta p^2}\right)p,\\
\Omega(p) &=& \frac{\sqrt{2}}{\pi}\left(\frac{\sqrt{\beta p^2}}{\sqrt{1+2\beta p^2}-1}\right)sin\left[\frac{\sqrt{2}\pi\left(\sqrt{1+2\beta p^2}-1\right)}{2\sqrt{\beta p^2}}\right].
\end{eqnarray}

Frassino and Panella used the concept of minimal length and the generalised uncertainty principle to obtain the finite energy between the uncharged plane plates. They explicitly derived the Hamiltonian and the corrections to the Casimir energy due to the minimal length. Up to a first order correction term in the minimal uncertainty parameter $\beta$, the Casimir energy for the two different cases of construction of maximally localized states are obtained as \cite{Frassino12}
\begin{equation}
E_{i}=-\frac{\pi^2}{720 }\frac{S}{a^3}\left[1+\xi_{i}\frac{\beta}{a^2}\right],
\end{equation}
where
\begin{eqnarray}
\xi_{KMM} &=& \pi^2\left(\frac{28+3\sqrt{10}}{14}\right),\\
\xi_{DGS} &=& 4\pi^2\left(\frac{3+\pi^2}{21}\right).
\end{eqnarray}

The corresponding energy density and pressure become
\begin{eqnarray}
\rho_i(a) &=& -\frac{\pi^2}{720 }\frac{1}{a^4}\left[1+\xi_{i}\frac{\beta}{a^2}\right],\\
p(a)      &=& -\frac{\pi^2}{240 }\frac{1}{a^4}\left[1+\frac{5}{3}\xi_{i}\frac{\beta}{a^2}\right],
\end{eqnarray}
where $i$ represents either the KMM construction or the DGS construction. 

\section{GUP corrected Casimir wormholes}
The plate separation distance $a$ may now be replaced by the radial coordinate $r$ to obtain the GUP corrected Casimir energy density that may be considered as the energy density of the exotic matter of the wormhole.  Solving the wormhole field equations in an extended gravity, we obtain the shape function as
\begin{equation}
b_i(r,\beta)=r_0+b_1\left(\frac{1}{r}-\frac{1}{r_0}\right)+\frac{b_1\xi_i\beta}{3}\left(\frac{1}{r^3}-\frac{1}{r_0^3}\right).
\end{equation}
The third term is the correction term due to the generalized uncertainty principle. Here the subscript $i$ denotes either the KMM construction or the DGS construction. The GUP correction term is proportional to the minimal uncertainty parameter $\beta$. Clearly in the limit $\beta\rightarrow 0$, the shape function reduces to that of the Casimir wormhole. These shape functions obviously satisfy the throat condition and the flare out condition. In Figures 3(a) and 3(b) the shape functions of the GUP corrected Casimir wormholes are shown for three different values of $\lambda$. The two types of construction of the maximally localized states are considered in the figures. We have considered $\beta=0.1$ and $r_0=1$ in drawing the figures. Since the two constructions differ only in the value of $\xi_i$, obviously the same has been reflected in the figures. In fact, we have $\frac{\xi_{KMM}}{\xi_{DGS}}=1.0923$ and therefore the contribution coming from the GUP correction term decreases by a factor of $1.0923$ for the DGS construction as compared to that of KMM construction. This leads to an increase in the value of shape function of DGS construction as compared to that of the KMM construction.

\begin{figure}[h!]
\begin{center}
\includegraphics[scale=0.4]{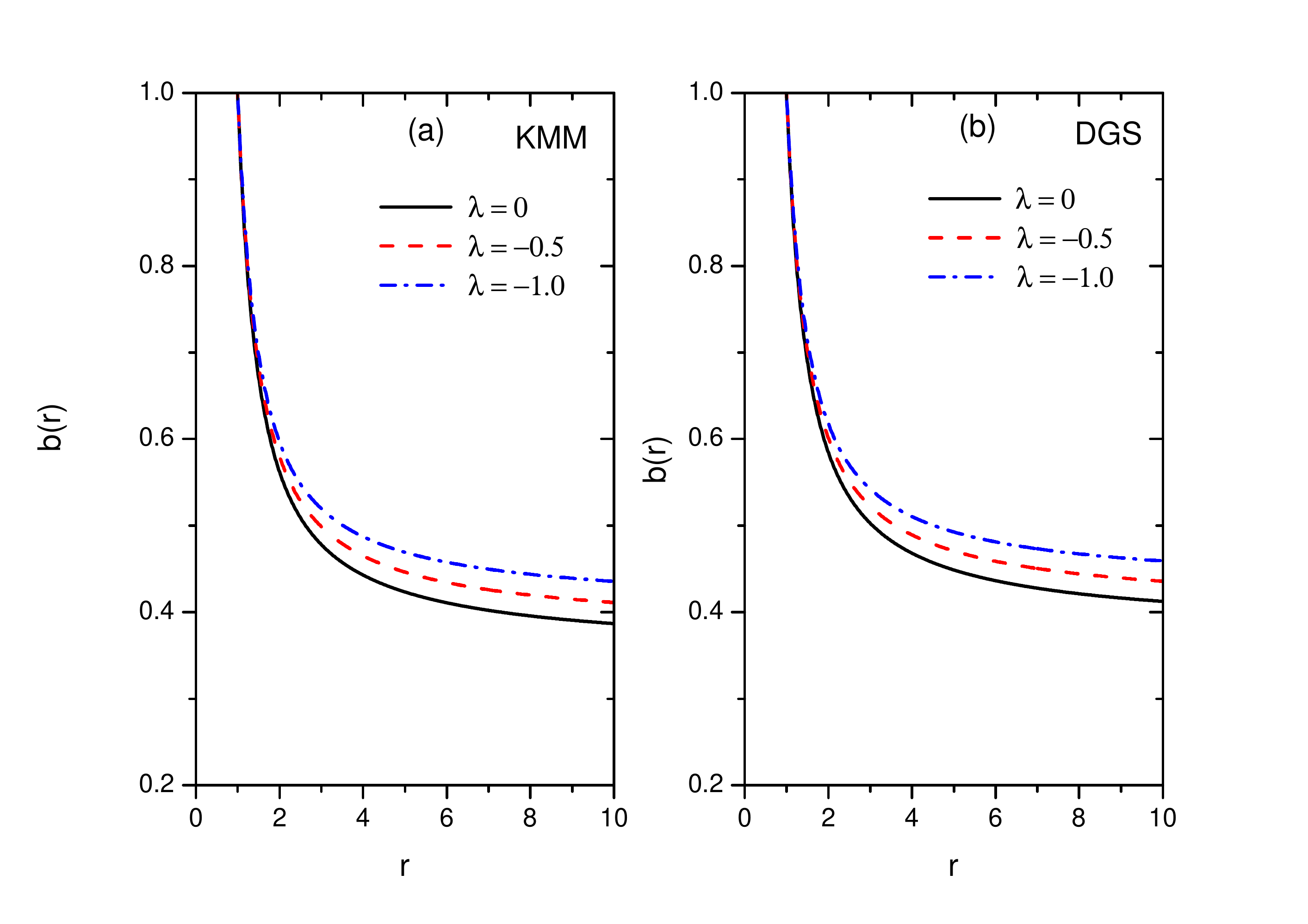}
\caption{(a) Shape function $b(r)$ of the GUP corrected Casimir wormhole for three different values of $\lambda$ for KMM construction. (b) The shape function for DGS construction. We have taken $r_0=1$ and $\beta = 0.1$. In the figure $\lambda=0$ corresponds to that in GR. }
\end{center}
\end{figure}

\begin{figure}[h!]
\begin{center}
\includegraphics[scale=0.4]{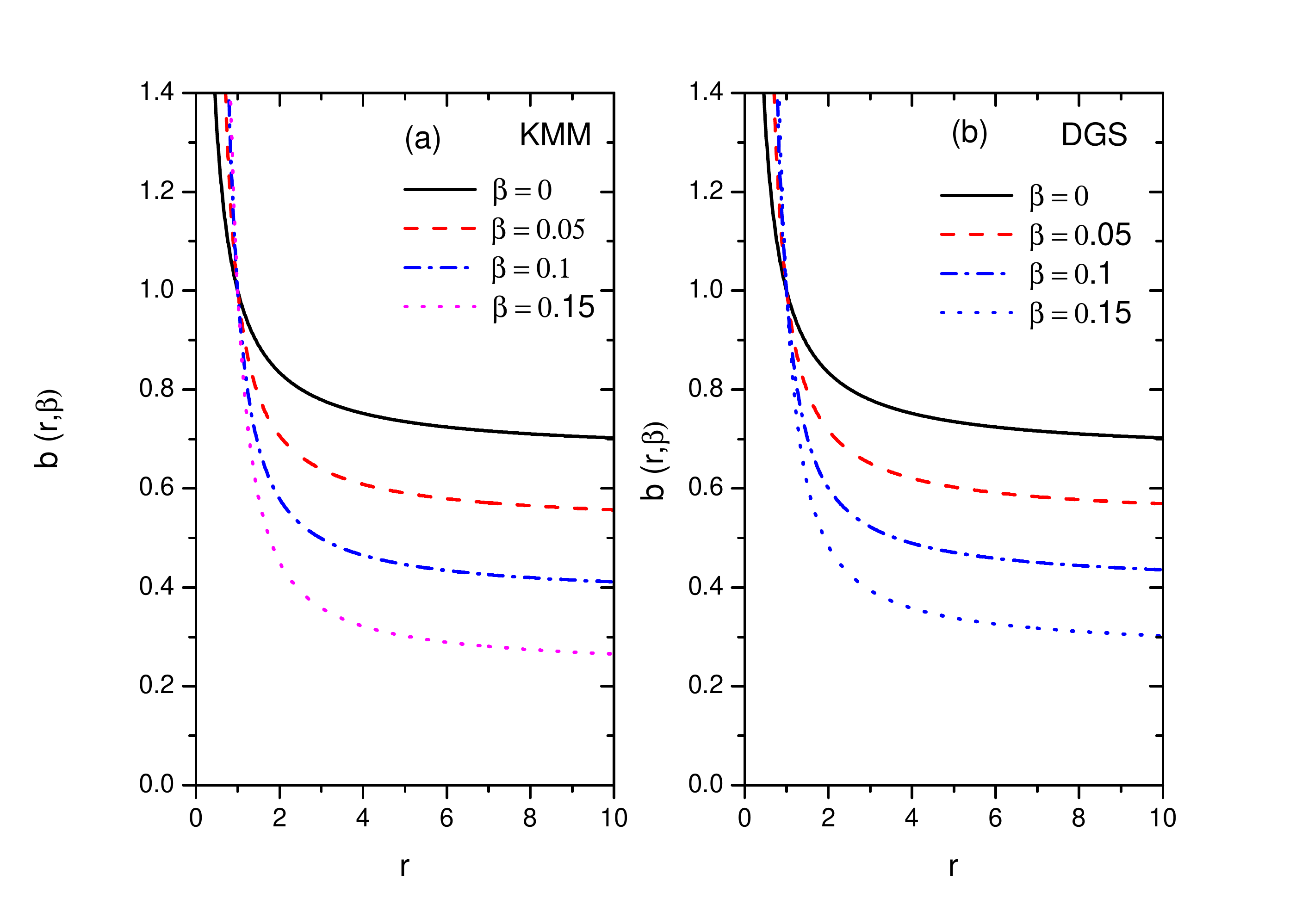}
\caption{(a) Shape function $b(r)$ of the GUP corrected Casimir wormhole for four different values of $\beta$ for KMM construction. (b) Shape function for DGS construction. We have taken $r_0=1$ and $\lambda = -0.5$. In the figures $\beta=0$ corresponds to the Casimir wormhole without GUP correction. }
\end{center}
\end{figure}

In order to asses the effect of the GUP correction on the shape function, we have plotted $b(r)$ for different values of $\beta$ in Figures 4(a) and (b). Since the GUP correction term becomes a negative contribution beyond the wormhole throat and is proportional to the minimal length parameter $\beta$, an increase in the value of $\beta$ obviously decreases the value of the shape function. 

The radial pressure and tangential pressure for the GUP corrected Casimir wormhole become
\begin{eqnarray}
p_r(r,\beta) &=& -\kappa^{\prime}_T\frac{1}{3r^6r_0^3}\left[3r^3r_0^4+3b_1r^2r_0^2(r_0-r)+b_1\xi_i\beta\left(r_0^3-r^3\right)\right],\\
p_t(r,\beta) &=& \kappa^{\prime}_T\frac{1}{6r^6r_0^3}\left[3r^3r_0^4+3b_1r^2r_0^2(2r_0-r)+b_1\xi_i\beta\left(2r_0^3-r^3\right)\right].
\end{eqnarray}

From the expressions of the radial and tangential pressure we obtain the corresponding EoS parameters as
\begin{eqnarray}
\omega_r(r,\beta) &=& \frac{1}{3b_1r_0^3}\left[\frac{3r^3r_0^4+3b_1r^2r_0^2\left(r_0-r\right)+b_1\xi_i\beta\left(r_0^3-r^3\right)}{\xi_i\beta+r^2}\right],
\end{eqnarray}
and
\begin{eqnarray}
\omega_t(r,\beta) &=& -\frac{1}{6b_1r_0^3}\left[\frac{3r^3r_0^4+3b_1r^2r_0^2\left(2r_0-r\right)+b_1\xi_i\beta\left(2r_0^3-r^3\right)}{\xi_i\beta+r^2}\right]\\
                  &=& -\frac{1}{2}\left[\omega_r(r,\beta)+1-\frac{\xi_i\beta}{\xi_i\beta+r^2}\right].
\end{eqnarray}

The anisotropy in the pressure of the exotic matter of the wormhole becomes
\begin{equation}
\triangle \omega (r,\beta)= \frac{\omega_t(r,\beta)}{\omega_r(r,\beta)}=-\frac{1}{2}-\frac{1}{2\omega_r(r,\beta)}\left[1-\frac{\xi_i\beta}{\xi_i\beta+r^2}\right],
\end{equation}
which reduces to 
\begin{equation}
|\triangle \omega (r_0,\beta)|=\frac{1}{2}\left[1+\frac{b_1}{r_0^2}\right].
\end{equation}
It is interesting to note that, the GUP modification to the Casimir energy affects the wormhole pressure both in the radial and tangential directions but at the throat, the magnitude of anisotropy in the pressure remains the same as that without modification.

The pressure anisotropy for the GUP corrected Casimir wormholes is shown for different values of the GUP parameter $\beta$ in Figures 5(a) and (b). Beyond the wormhole radius, the anisotropy factor decreases with the increase in $\beta$. However, for a radial distance less than the throat, $\triangle \omega (r,\beta)$ increases with $\beta$. In comparison to the usual Casimir wormholes, the behaviour of $\triangle \omega (r,\beta)$ is quite different  at a radial distance $r<r_0$. At the wormhole throat, the pressure anisotropy parameter becomes independent of the GUP correction parameter $\beta$.
\begin{figure}[h!]
\begin{center}
\includegraphics[scale=0.4]{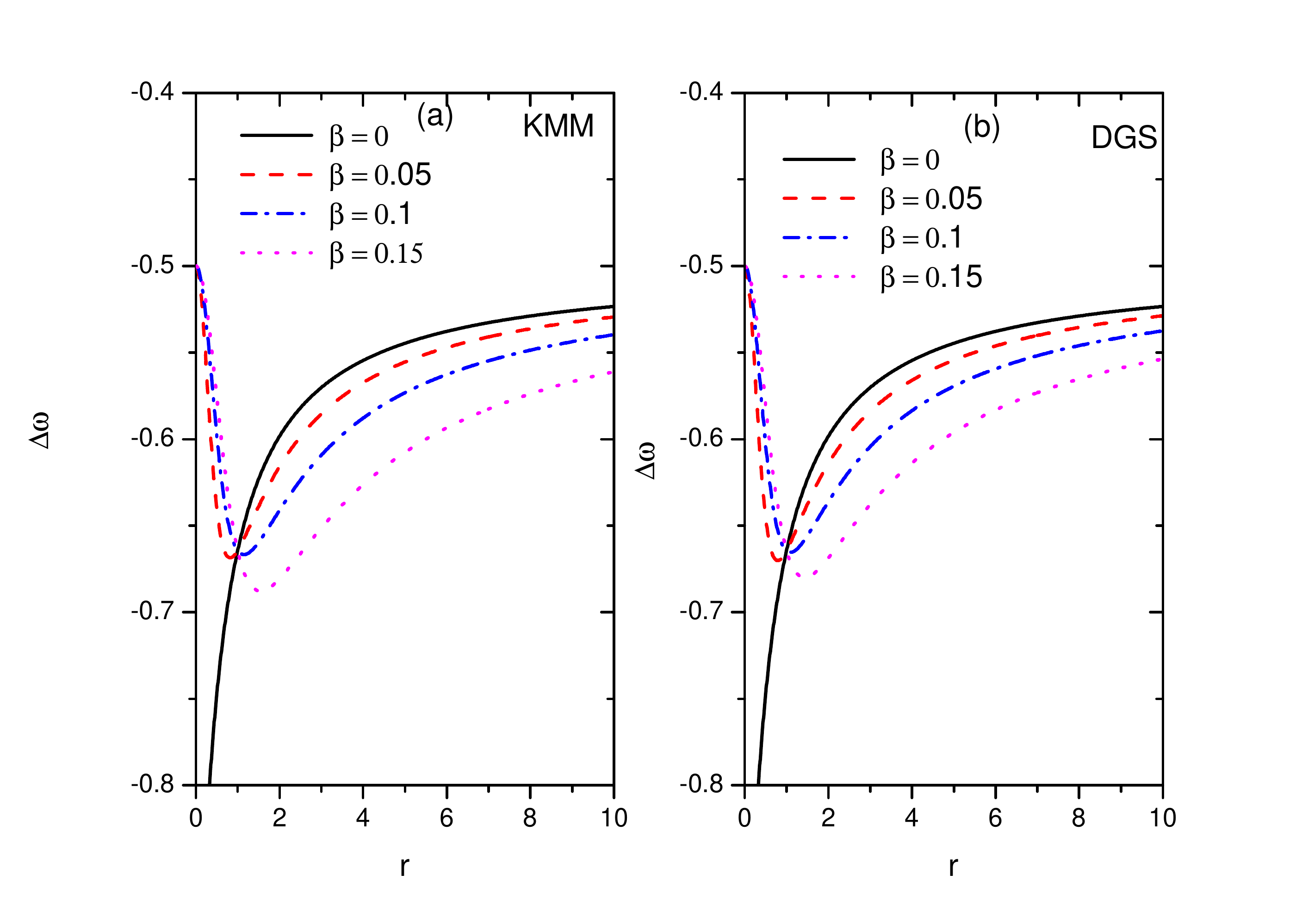}
\caption{ $\triangle \omega(r,\beta)$ of the GUP corrected Casimir wormhole for four different values of $\beta$. We have taken $r_0=1$ and $\lambda = -0.5$. In the figure $\beta=0$ corresponds to the Casimir wormhole without GUP correction. }
\end{center}
\end{figure}

\section{Energy conditions}
In general, due to the presence of exotic matter in wormholes, some energy conditions are violated. Particularly the NEC defined as $t_{\mu\nu}k^{\mu}k^{\nu}\geq 0$ or $\rho(r,\beta)+p_r(r,\beta)\geq 0$ is violated. In this section, we intend to check whether the GUP corrected Casimir wormholes in the extended gravity theory satisfies the energy conditions.
The NEC for the GUP corrected Casimir wormholes can be assessed from the expression

\begin{eqnarray}
&\textbf{NEC}:&~~~~~~  \rho(r,\beta)+p_r(r,\beta)= -\frac{\kappa^{\prime}_T}{r^3}\left[r_0+\frac{b_1}{r}\left(\frac{2r_0-r}{r_0}\right)+\frac{\xi_i\beta b_1}{r^3}\left(\frac{4r_0^3-r^3}{3r_0^3}\right)\right].\label{eq:55}
\end{eqnarray}

In the above, the GUP correction term is proportional to the minimal uncertainty parameter $\beta$.  For a radial distance $r<r_0$, obviously the right hand side of eqn. \eqref{eq:55} is a negative quantity and therefore the NEC is violated. With an increase in $\beta$, the contribution becomes more and more negative. The role of $\lambda$ is to minimise the NEC violation through the factor $b_1$. In the limit $\beta \rightarrow 0$, eqn. \eqref{eq:55} reduces that of a Casimir wormhole:
\begin{equation}
\rho(r,\beta)+p_r(r,\beta)= -\frac{\kappa^{\prime}_T}{r^3}\left[r_0+\frac{b_1}{r}\left(\frac{2r_0-r}{r_0}\right)\right].
\end{equation}

At the throat, this expression \eqref{eq:55} reduces to
\begin{equation}
\rho(r_0,\beta)+p_r(r_0,\beta)= -\frac{\kappa^{\prime}_T}{r_0^3}\left[r_0+\frac{b_1}{r_0}+\frac{\xi_i\beta b_1}{r_0^3}\right].\label{eq:NEC}
\end{equation}
Since the right side of the eq.\eqref{eq:NEC} is a negative quantity, it is obvious that, the NEC is violated by the GUP corrected Casimir wormhole at the throat.

\begin{figure}[h!]
\begin{center}
\includegraphics[scale=0.65]{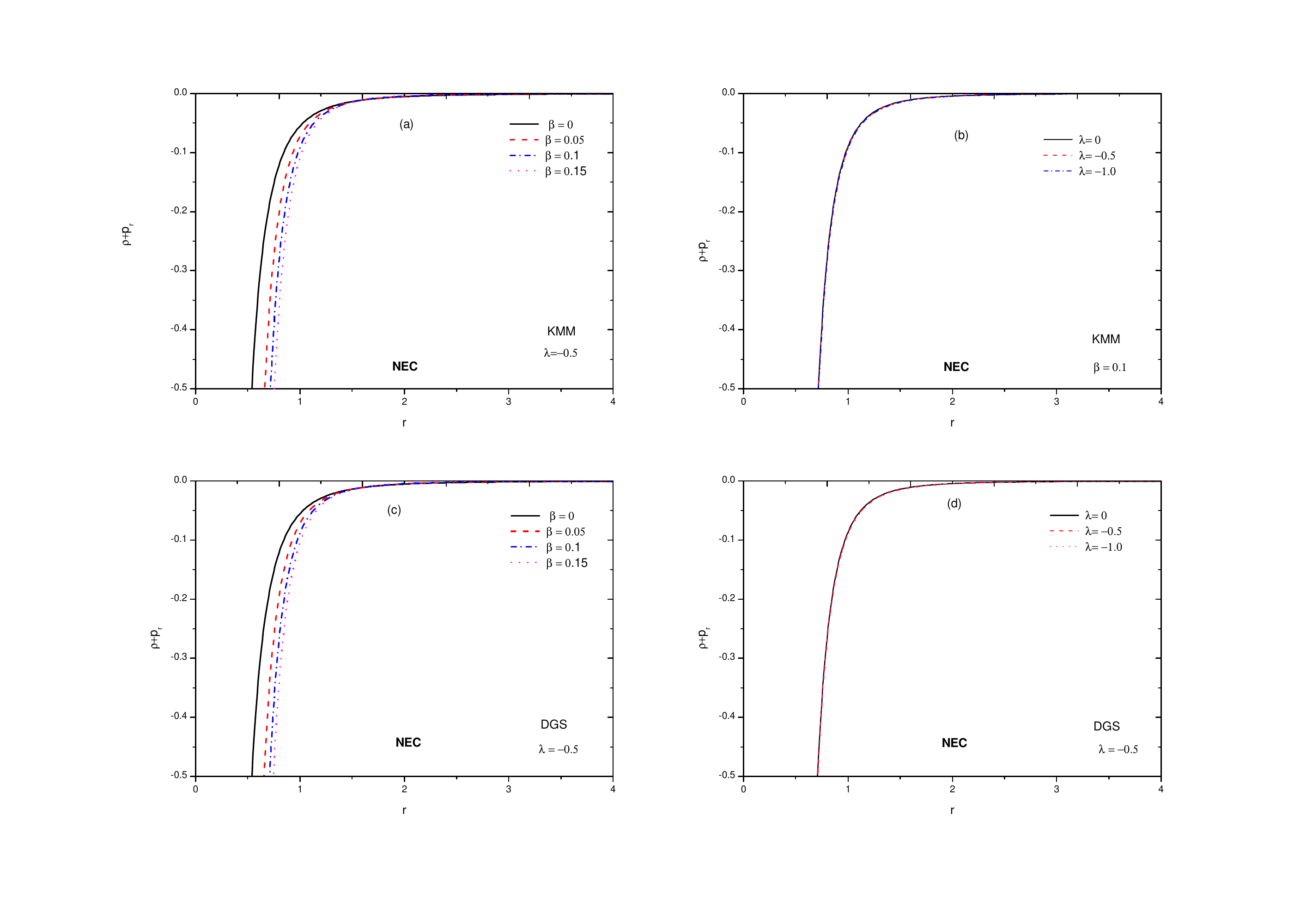}
\caption{ (a)\textbf{NEC:} GUP corrected wormhole in KMM construction for different $\beta$. (b)\textbf{NEC:} GUP corrected wormhole in KMM construction for different $\lambda$. (c)\textbf{NEC:} GUP corrected wormhole in DGS construction for different $\beta$. (d)\textbf{NEC:} GUP corrected wormhole in DGS construction for different $\lambda$. We have taken $r_0=1$. In the figures $\beta=0$ corresponds to the Casimir without GUP correction. }
\end{center}
\end{figure}
The strong energy condition(SEC) is given by $\rho(r, \beta)+2p_t(r,\beta)\geq 0$. From the expressions of the Casimir energy density and the radial pressure we have 
\begin{equation}
\textbf{SEC}:~~~~~~\rho(r, \beta)+2p_t(r,\beta)=\frac{\kappa^{\prime}_T}{r^3}\left[r_0-\frac{b_1}{r}+\frac{b_1(2r_0-r)}{rr_0}+\frac{\xi_i\beta  b_1}{r^3}\left(\frac{2r_0^3-r^3}{3r_0^3}-1\right)\right],
\end{equation}
which reduces to
\begin{equation}
\rho(r_0,\beta)+2p_t(r_0,\beta)=\frac{\kappa^{\prime}_T}{r_0^3}\left[r_0-\frac{2\xi_i\beta b_1}{3r_0^3}\right]
\end{equation}
at the throat. A violation of SEC at the throat requires that $r_0<\left(\frac{2\xi_i\beta b_1}{3}\right)^{1/4}$.

Another way to express the SEC is $\rho(r,\beta)+p_r(r,\beta)+2p_t(r,\beta) \geq 0$. For this statement, we have
\begin{equation}
\rho(r,\beta)+p_r(r,\beta)+2p_t(r,\beta)=-\frac{2}{3}\frac{\kappa^{\prime}_T \xi_i\beta  b_1}{r^6}.
\end{equation}
Clearly, this condition is violated by the GUP corrected Casimir wormhole. In Figures 6(a)-6(d), the NEC for the GUP corrected Casimir wormholes are shown for the two specific constructions of maximally localized states. In the figure 6(a), NEC for KMM construction is plotted for different values of $\beta$.  The GUP corrected Casimir wormhole violates the NEC for all the range of radial distances considered in the work. An increase in the GUP parameter $\beta$ makes the  energy condition more negative for $r<r_0$. However, after a radial distance $\frac{r}{r_0}>1.6$, an increase in $\beta$ makes the energy condition less negative. The effect of $\lambda$ on the energy condition of KMM construction is shown in figure 6(b). The effect of $\lambda$ is not visible in the figures. However, a decrease in the value of $\lambda$ decreases the value of $b_1$ and thereby it minimises the energy condition violation. Concerning the effect of $\beta$ and $\lambda$ similar behaviour have been observed for the DGS construction. 
\subsection{Exotic matter content}
Traversable wormholes with exotic matter content violate the average null energy condition (ANEC) \cite{Morris88, Hochberg98}.  Since quantum effects induce some energy condition violation \cite{Epstein65}, it is  pertinent to think of how much ANEC violating matter is present in the spacetime. Visser et al. have proposed a volume integral theorem that quantifies the amount of ANEC violating matter present in the spacetime \cite{Visser03}.

Using the volume integral theorem of Visser et al. we calculate the exotic matter content of the GUP corrected wormholes violating the ANEC as
\begin{eqnarray}
m=\oint \left[\rho(r,\beta)+p_r(r,\beta)\right]dV.
\end{eqnarray}
Since $\oint dV=2\int\limits_{r_0}^{\infty} dV=8\pi\int\limits_{r_0}^{\infty} r^2dr$, we should evaluate the integral

\begin{equation}
m=-8\pi\kappa^{\prime}_T\int\limits_{r_0}^{\infty}  ~\frac{1}{r}\left[r_0+\frac{b_1}{r}\left(1+\frac{(r_0-r)}{r_0}\right)+\frac{\xi_i\beta b_1}{r^3}\left(1+\frac{r_0^3-r^3}{3r_0^3}\right)\right]~dr.
\end{equation}

However, it would be useful, if we have a wormhole whose field only deviates from that of Schwartzchild in the region from the throat to a certain radius $\mathcal{R}$, then we have
\begin{eqnarray}
m &=& -8\pi\kappa^{\prime}_T\int\limits_{r_0}^{\mathcal{R}}  ~\frac{1}{r}\left[r_0+\frac{b_1}{r}\left(1+\frac{(r_0-r)}{r_0}\right)+\frac{\xi_i\beta b_1}{r^3}\left(1+\frac{r_0^3-r^3}{3r_0^3}\right)\right]~dr,\\
   &=& 8\pi\kappa^{\prime}_T\left[\left(\frac{b_1}{r_0}-r_0+\frac{\xi_i\beta b_1}{3r_0^3}\right)~ln\left(\frac{\mathcal{R}}{r_0}\right)+\frac{4\xi_i\beta b_1}{9}\left(\frac{1}{\mathcal{R}^3}-\frac{1}{r_0^3}\right)\right].
\end{eqnarray}
Obviously for a region confined to the throat, i.e. $\mathcal{R}=r_0$, no exotic matter is required to support the wormhole.
However, for a region outside the throat, we may consider $\mathcal{R}=r_0+\epsilon$, so that 
\begin{eqnarray}
m= 8\pi\kappa^{\prime}_T\left[\left(\frac{b_1}{r_0}-r_0+\frac{\xi_i\beta b_1}{3r_0^3}\right)~ln\left(1+\frac{\epsilon}{r_0}\right)+\frac{4\xi_i\beta b_1}{9\mathcal{R}^3}\left(1-\left(1+\frac{\epsilon}{r_0}\right)^3\right)\right],
\end{eqnarray}
which in the limit $\frac{\epsilon}{r_0} \ll 1$ reduces to
\begin{eqnarray}
m\simeq -\frac{4\xi_i\beta b_1}{3\mathcal{R}^3}\frac{\epsilon}{r_0}.
\end{eqnarray}
In the limit of General  Relativity, we may have
\begin{eqnarray}
m= -\frac{2}{135}\frac{\pi^3\xi_i\beta}{\mathcal{R}^3}\frac{\epsilon}{r_0}.
\end{eqnarray}
In other words, a small amount of exotic matter is required to support a traversable wormhole in a region close to the throat. In fact, the total amount of ANEC violating matter can be reduced by considering suitable wormhole geometry. 

\section{Conclusion}

In this paper, we have modelled traversable static wormholes in the framework of an extended gravity theory. In this modified gravity theory, the geometrical action contains a term proportional to the trace of the energy momentum tensor. This additional term provides an explanation for the late time cosmic speed up issue without the need of any exotic degrees of freedom in the matter field. Such a theory may be associated with the existence of imperfect fluids and has been motivated from the quantum effects of particle production. It is needless to mention here that, the extended gravity theory has been widely investigated in literature concerned with many issues in cosmology and astrophysics.

Casimir effect that appears due to distorted quantized field of the vacuum between two plane parallel plates is known to be associated with exotic pressure and energy which may be realisable in the laboratory. Obviously, such exotic matter violates the energy conditions. Since wormholes are solutions of Einstein field equations and contain exotic matter and violate the null energy condition, the quantum nature of Casimir effect may be useful for modelling these exotic objects. In the present work, we have modelled such traversable wormholes with exotic matter content favouring Casimir effect in the framework of the extended gravity theory. From the Casimir energy density, we have integrated the extended gravity field equations to obtain the shape function of the wormhole metric. The obtained shape function satisfies the throat condition and the flare out condition well. The slope of the shape function vanishes asymptotically. Also, we have observed the effect of the modified gravity on the shape function. At a radial distance away from the wormhole throat, a decrease in the value of the extended gravity parameter results in an increase in the shape function. However, at a radial distance closer to the throat, the effect of the extended gravity is not substantial. We have defined radial and tangential equation of state parameters as the ratio of the respective pressure to the Casimir energy density. While the radial equation of state parameter is obtained to be an increasing function of the radial distance, the tangential equation of state parameter is a decreasing function. The extended gravity parameter also affects the behaviour of the radial and tangential equation of state parameters to a large extent at least at distances away from the throat. The pressure anisotropy defined as the ratio of  the tangential to radial equation of state parameter decreases with a decrease in the parameter $\lambda$.

The concept of minimal length in quantum gravity theories has led to the proposal of generalised uncertainty principle. In the present work, we have explored the effect of GUP on the traversable wormholes with exotic Casimir energy density. Since there can be many ways to construct the maximally localized quasi quantum states using different possible generic functions, there can be ample ways to model the Casimir wormholes. However, in the present work, we have considered two well known construction of the quasi quantum states. Using these KMM and DGS construction, we have modelled the Casimir wormholes in the framework of the extended gravity theory. The shape function of the GUP corrected Casimir wormhole assumes a lower value than the usual Casimir wormhole at a radial distance away from the throat. However, it assumes a higher value at a distance below the throat. It has been observed that, the GUP correction has substantially changed the dynamics of the pressure anisotropy of the Casimir wormhole. We have also investigated the violation of the energy conditions for the GUP corrected wormhole. The GUP corrected Casimir wormholes violate the energy condition more evidently with an increase in the minimal length parameter for both the types of construction. Even though the effect of  the extended gravity parameter is not visible, it has a role to minimise the  energy condition violation. Also, we have calculated the exotic matter content of the wormhole using the volume integral theorem of Visser et al. \cite{Visser03} and found that a small amount of exotic matter is required to support the wormhole.

We may remark here that, lensing properties of wormholes lead to their possible detection which may be used to test different gravity theories. Another aspect is the calculation of the deflection angle of light from wormhole at least at the weak field limit. The lensing data and deflection angles from observations may  be used to constrain the modified gravity parameters. In some recent works, Dai and his coworkers have suggested that the gravitational perturbations in the space on the other side of the wormhole should affect the trajectories of the objects in the vicinity of a wormhole \cite{Dai2019, Simonetti2020}. Looking for the effect of a perturbing object orbiting in one side of the wormhole on the orbit of a star at the other side may be a more sensitive test of observing wormholes \cite{Simonetti2020}. The model presented in the present work may also be confronted with observational lensing data that may come up. In the present work, we have used a simple extended gravity theory to explore the physical behaviour of the Casimir wormholes and GUP corrected wormholes. One may use any other extension of the functional $f(T)$ else than the linear one to obtain more general wormholes.

\section*{Acknowledgement}
SKT thanks IUCAA, Pune (India) for providing support during an academic visit where a part of this work is carried out.

\end{document}